# Co-Design of Memory-Storage Systems for Workload Awareness with Interpretable Models

Jay Sarkar*, Vamsi Pavan Rayaprolu*, Abhijeet Bhalerao*
Micron Technology
San Jose, California, U.S.A.
{jaysarkar, vrayaprolu, abhalerao}@micron.com

*Abstract*— Solid-state storage architectures based on NAND or emerging memory devices (SSD), are fundamentally architected and optimized for both reliability and performance. Achieving these simultaneous goals requires co-design of memory components with firmware-architected Error Management (EM) algorithms for density- and performance-scaled memory technologies. We describe a Machine Learning (ML) for systems methodology and modeling for co-designing the EM subsystem together with the natural variance inherent to scaled silicon process of memory components underlying SSD technology. The modeling analyzes NAND memory components and EM algorithms interacting with comprehensive suite of synthetic (stress-focused and JEDEC) and emulation (YCSB and similar) workloads across Flash Translation abstraction layers, by leveraging a statistically interpretable and intuitively explainable ML algorithm. The generalizable co-design framework evaluates several thousand datacenter SSDs spanning multiple generations of memory and storage technology. Consequently, the modeling framework enables continuous, holistic, data-driven design towards generational architectural advancements. We additionally demonstrate that the framework enables Representation Learning of the EM-workload domain for enhancement of the architectural design-space across broad spectrum of workloads.

## I. INTRODUCTION AND MOTIVATION

SSD technology has become mainstream across cloud datacenters, edge and mobile applications with increasing scale and sophistication. As SSD technology enables capabilities to accommodate storage workloads spanning hot, warm and colder data across the system stack of datacenters, there is also continuing research and expectation for architectural optimization for SSD technology towards adaptation for increasingly complex and evolving capabilities [1], [2]. At the same time, increasingly sophisticated NAND memory technologies underlie the storage architecture [2]. Therefore, increasing sophistication of EM architecture is necessary to manage bit-errors and defects naturally arising from NAND memory components, to meet and exceed competitive, strict and standards-defined qualification of reliability error rates and performance at the host interface, across SSD populations of several tens of million or more in production deployment [3], [5]. This poses a significant challenge for designing the EM stack of the storage architecture, to ensure that NAND bit-errors and defects are adaptively tolerated and managed by the EM subsystem - while providing direct feedback and flexibility for deciding the threshold of architectural acceptability of bit-errors and defects statistically evident across large SSD populations undergoing diverse workloads [3]. Furthermore, when silicon capabilities are challenged by increasing complexity of scaled technologies, it becomes necessary and fruitful to innovatively evolve the EM architecture, towards holistically integrated storage system-level intrinsic and extrinsic reliability.

Significant motivation therefore exists for data-driven co-design of the interface of the EM architecture with the silicon memory components for optimal system-level benefits [6]. Sampling with small datasets is almost never effective to achieve such a goal, due to unavoidable selection bias inherent in small datasets. Hence, we address this problem with largescale data-driven co-design, where the effect of diverse, heterogenous workloads on large SSD populations can be leveraged by storage architects in an iterative feedback loop. To enable co-design, we leveraged an ML-for-Systems modeling framework as the "analytical lens" applied to modeling the EM architecture in concert with measured memory silicon behavior across diverse, heterogenous workloads. Given the plurality of memory components in every SSD, and natural variance of bit-errors and defects inherent to, and across those memory components, an ML-for-Systems approach proves essential and fruitful [4], [32]. To that end, we consciously adopt our modeling framework to be based on a carefully selected interpretable ML model that is intuitively understandable and explainable to storage architects, in contrast to black-box models such as based on neural networks [3], [4]. Such a design choice for the framework specifically ensures it is transparently understood, trusted and accepted in production deployment. In addition to addressing the statistical diversity inherent in memory component parameters, and ensuring ML modeling interpretability necessary for system architects, the framework in this context also needs to be generalizable across multiple, evolving memory and SSD technologies [4]. This aspect ensures that the once-developed framework only requires minimal revisions or maintenance and does not require re-development for each new SSD or memory technology generation. It is therefore with consideration of such necessities and associated design principles; we enable the co-design framework for applications across multiple generations of storage and memory technology.

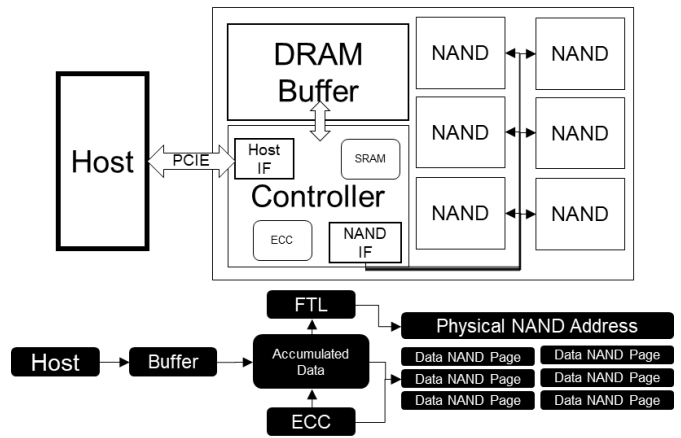

Figure 1. SSD architecture incorporating firmware-architected EM with associated controller architecture and subsystems, interfacing with host server.

---

*Authors contributed equally to this work. © Authors.

## II. Background and Related Research

### A. Solid-State Storage Workloads

Storage workloads have evolved significantly, from enterprise systems to datacenters, and such evolution is especially relevant currently for Artificial Intelligence applications in warehouse-scale computing infrastructures [8]. For example, for datacenter applications, new SSD architectures have been developed in the recent past, to address certain classes of datacenter workloads, such as the Zoned Namespace (ZNS) architecture [9]. To optimally encompass a broad-spectrum of evolving workloads, storage architectures therefore often need to be empirically evaluated and validated against a gamut of workloads that assess their robustness and performance. The classification of SSD workloads into real workloads and synthetic workloads is to achieve optimal architectural choices for both NAND memory and SSDs in terms of both reliability and performance [10]. Real workloads are generated through trace capture from actual production environments, providing authentic I/O patterns observed during real-world application use [11]. The SNIA Real World Storage Workload (RWSW) Performance Test Specification describes real-world storage workload I/O capture, characterization, and testing methodologies [12]. Synthetic workloads represent artificially generated patterns preserving key properties of realistic workloads, including request interarrival time distributions, request size distributions, operation mixes, and locality patterns [13]. Advanced synthetic workload generation may employ sophisticated modeling techniques to accurately recreate I/O loads of datacenter applications [14].

### B. Industry Standard JEDEC Workloads

The JEDEC-218 specification provides fundamental guidelines for SSD technology to ensure lifetime and reliability, while JEDEC-219 defines standardized endurance workloads for both client and enterprise application classes [15]. These standards establish critical specifications such as maximum allowable Functional Failure Rates and Uncorrectable Bit Error Rates (UBER) [16]. They are also a critically necessary class of workloads for production-grade SSD qualification, adopted by producers and consumers of SSD technology as well-established industry standards for over a decade [16], [17]. These standards follow physics- and architecture-aligned workload design to specifically stress memory silicon reliability, Flash Translation Layer and associated interactions with host filesystem and applications, across the host-interface stack.

### C. YCSB Standardized Benchmarking

The Yahoo! Cloud Serving Benchmark (YCSB) exemplifies systematic synthetic workload generation, providing standardized database application workload patterns [18]. YCSB enables parameterized workload definitions including configurable read/write ratios, access patterns, and variable record sizes. The framework supports more than 40 NoSQL databases and provides essential performance metrics including throughput, latency, and response time distributions.

### D. Proprietary Workloads

Proprietary workloads are derived from Design Failure Mode and Effects Analysis methodologies and design-specific considerations [19]. These typically include firmware scenario tests and stresses, thermal throttling stresses, wear-leveling algorithm scenario testing, and garbage collection efficacy tests. For 3D NAND memory, proprietary workloads may address unique characteristics including layer-to-layer process variation, early retention loss, and retention interference [20]. Certain proprietary workloads may also address extreme operating conditions that can occur very rarely in deployed systems. Modern stress testing may also incorporate specialized patterns such as Row Hammer testing, initially relevant for DRAM technology, but adapted for storage technology for phenomenologically similar cell-to-cell interference mechanisms [21]. Such aspects represent the evolution of qualification methodologies to address emerging challenges in continually scaled memory and storage technology.

Thus, a comprehensive suite of workloads provides the foundation for methodically ensuring SSD robustness and performance across diverse operational scenarios, towards high-volume deployments in evolving computing environments. One of our modeling framework goals has been to derive unified and holistic understanding of the impact of the entire suite of workloads on the storage and EM architecture both in current state as well during future evolutions.

### E. Error Management Architecture

The benefits of lower latency and higher data throughput from SSD technology and lower Total-Cost-of-Ownership (TCO) relative to hard-disk technologies have led to continuous expansion of storage workloads provisioned to SSD technology in datacenters. Fundamental to SSD technology is the sophistication of EM schemes and algorithms that enable the expanding workload spectrum [6]. SSD technology is inherently architected in a distributed manner, to exploit maximum robustness and performance achievable from the parallelism of internally distributed data layout. The distributed architecture for internal data layout, along with firmware-architected EM schemes, are described well by Cai et al. [22].

The memory hardware architecture internal to SSD designs is reproduced in Fig. 2 for contextual completeness, wherein units of host data are shown internally arranged and aligned to purposefully span hierarchical units of pages, superpages, blocks, superblocks, dies, chips and channels. Over the recent several years, the performance landscape of datacenter SSD technology has evolved dramatically. Latency expectations have become increasingly stringent, with Cloud applications often requiring microsecond-level response times with high consistency, often targeting tail latencies at 99.9th to 99.999th percentile and can manifest as extrinsic "fail-slow" reliability issues in datacenters [23, 30]. A significant contributor to SSD latency, beyond read collisions at high queue depths, is the overhead introduced by EM operations. Optimizing EM is therefore an evolving research area, driven by the scaling of memory technology and increasing density of bits per cell, where known NAND mechanisms such as data retention and read disturb, and new challenges like lateral charge migration relevant to charge trap memory technology, are increasingly more relevant [24]. Research efforts include development of advanced error correction schemes and read-retry methodologies that dynamically adjust read voltage thresholds to align with the distribution of stored data in TLC or QLC NAND [25]. Advanced recovery techniques such as soft decoding and Neighbor-Cell Assisted Error Correction (NAC) can significantly extend EM overhead, posing challenges for latency-sensitive applications [26]. As technology scaling becomes more complex and latency expectations more demanding, careful orchestration of EM steps - both in terms of sequence and computational complexity - has become essential. Therefore, the pipelined sequence of EM steps, as illustrated in Fig. 3(a) needs to be carefully designed to balance the frequency of retry attempts vs. increasing latency observed by codewords as represented in Fig. 3(b) – towards improving the overall average throughput of codewords flowing through EM, while minimizing tail latency of the storage system. Analyzing the EM architecture thus yields rich insights into both workload characteristics and memory technology, offering dual perspectives on storage system reliability and performance. When carefully co-designed closely with memory technology and optimized to accommodate a broad range of workload and stress conditions — the EM architecture thus serves as a determinant of multiple figures of merit of the storage architecture. Research on EM has therefore often focused on increasing applications of data-driven, ML methods toward system design goals. Such imperatives and opportunities have motivated our research approach to system-level co-design with workload-awareness.

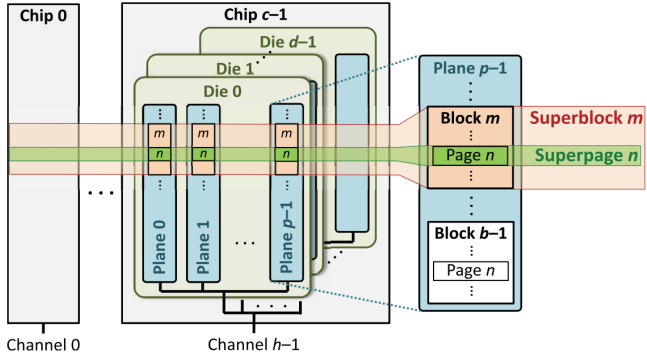

Figure 2. Internal memory organization in SSD technology. Reproduced from Cai et al [22], © IEEE 2017.

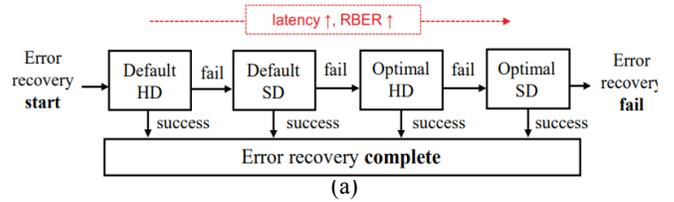

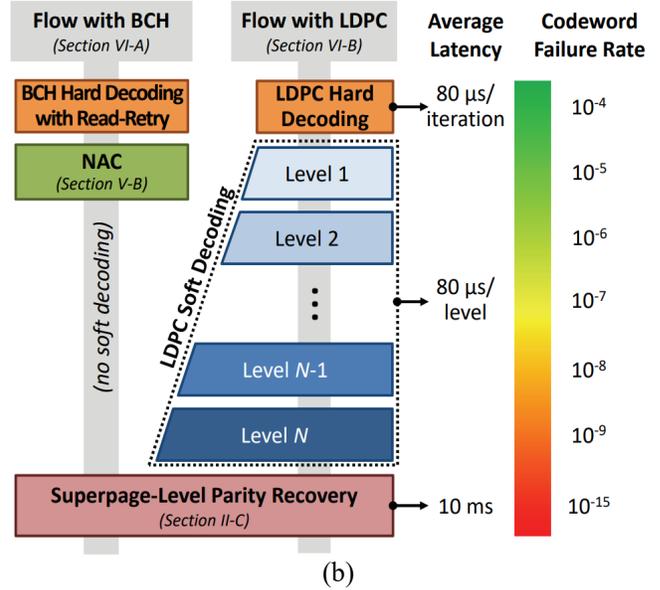

Figure 3. (a) A standard pipelined read-retry flow. Reproduced from Kang et al [27]. © IEEE 2020. (b) Example EM flow using BCH codes and LDPC codes and the corresponding average latency and codeword failure rate for each LDPC stage. Reproduced from Cai et al [22]. © IEEE 2017.

*F. Related Research*

In contrast to research on datacenter or Cloud storage systems from observational perspectives, which use external datasets (such as SSD or HDD S.M.A.R.T logs) available from the system and application stack, this research focuses on proactive design and optimization of SSD technology as a resilient and performant system of hardware and firmware [3], [5], [7]. In this research, the workload stimulus flows across the stack of filesystems and Flash Translation Layer abstractions, after initiation directly from the host workload and software applications. Thereafter, the response of the EM architecture to workload stimuli is measured and modeled as stimulated *internally* in the storage devices, towards noted co-design goals. Additionally, this research addresses the EM architecture with focus on holistic system design encompassing all memory components integrated with a controller -- and therefore differs from research on error correction code or NAND memory architecture that focuses on individual memory components, array or sub-array [2]. Our research may stimulate and influence similar research on the components and architecture of the EM stack, with the goal of optimization and innovation in concert with software applications at the host-level.

## III. WORKLOAD PROFILING CHARACTERIZATION DATASET

SSD population and profiling workloads evaluated in this work spans across three different datacenter-class SSD designs described in Table 1. While bearing standard NVMe datacenter-class storage architecture within specified lifetime of Program/Erase cycles [31], three SSD designs incorporated evolving memory silicon, EM with varying step counts, and spanning PCI Express generations, while being subjected to the gamut of workload classes described in Section II.

The population also represented an approximately Gaussian distribution of memory component counts per SSD, as described in Fig. 4. In all, approximately 0.5 million Triple-Level Cell (TLC) memory components were present in the overall population, with a broad distribution of associated SSD capacities, reflected by Fig. 4. The profiled population was therefore large and diverse, which captured significant silicon and system heterogeneity for the co-design goal.

Integer *counts* of activation for each step of EM needed by codewords, prior to successful error correction for reading out page-level data from the underlying memory architecture in Fig. 1-3, comprised the dataset obtained by continuous telemetry. Thus, our dataset reflected the effectiveness of the EM flow in reliable and complete retrieval of host and user data resident across NAND dies, despite bit-errors and defects from variety of mechanisms inherent in memory technology, while being stimulated by significantly heterogeneous workloads.

| SSD Generation | 3-D NAND by Layer Count | SSD Population ($n$) |
|---|---|---|
| SSD-A (PCIe Gen. $N$) | X | 906 |
| SSD-B (PCIe Gen. $N$+1) | X | 4027 |
| SSD-C (PCIe Gen. $N$+1) | Y | 3378 |

TABLE 1: WORKLOAD-PROFILED POPULATION OF 8311 SSDs, WITH DESIGN HETEROGENEITY SPANNING GENERATIONS OF MEMORY SILICON AND STORAGE ARCHITECTURE.

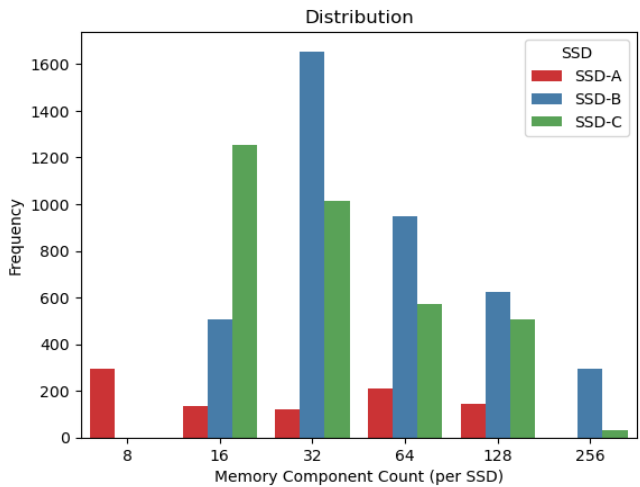

Figure 4. Frequency distribution representing ~0.5 million memory compoents underlying varying capacities of SSD that experience diverse, heterogenous profiling workloads described in Section II.

## IV. MODELING FRAMEWORK FOR CO-DESIGN

Given the large and heterogeneous resulting dataset comprising several thousand storage devices, it becomes exceedingly challenging, if not impossible, to bring forward comprehensive understanding with manual, non-algorithmic analysis. At the same time, the goal of co-design for storage architects excluded black-box algorithmic automation. This is because understanding and consequent reasoning across generational evolution of memory and EM, with architects in the loop, is central to carefully engineering an evolving EM stack. At the same time, it is also understood that historical memory or SSD generations do not offer automatic or labeled datasets that can be meaningful for a supervised ML framework – since the framework needs to continually evaluate new datasets from evolving and advancing generations of memory.

Given such necessities and constraints in the problem-space from a ML-for-Systems perspective, we adapted and developed from first principles an interpretable unsupervised ML modeling framework that algorithmically uses fundamental properties of Extreme-Value statistics, when appropriately mapped to EM algorithms and architecture. For the domain of EM system design, two key insights we leveraged are:

(a) Necessity of evolution of any EM step or aberrant silicon behavior would manifest as extreme values or tails in the behavioral distribution of the step's counts. Specifically, more (or, less) attempts for success for an EM step would be evident in the dataset, if that specific step is being overly necessary (or, not necessary at all) during the execution of profiling workloads, implying that the step's algorithm may be sub-optimal (or, unnecessary or ineffective).

(b) Simultaneously, storage architects would also need to holistically account for the full context of the EM architecture – such that one step's behavior in one SSD or one workload is not necessarily insightful. However, statistically consistent extreme-value behavior of any step or steps, evident across the large population of SSDs and memory components, for example, almost certainly indicates a necessity to re-evaluate that EM step and/or the associated memory silicon.

While parametric unsupervised ML models assume that the data underlying the modeling method belongs to a parametric distribution for fitting the data, a non-parametric modeling method does not make such an assumption. In this problem space spanning generations of storage technology, the specific nature of the distributions of the EM step activations are neither known, nor can be presumed *a priori*. Nor is it acceptable that the chosen ML model's multiple hyperparameters need to be tuned for every new generation of memory or storage. Hence, selecting a non-parametric, unsupervised ML model became imperative for generalizability and interpretability goals across generations of storage technology.

With the noted principled co-design goals in mind, we chose the Empirical Cumulative Distribution Functions (ECOD) as the deployed ML model [28]. The model empirically constructs the Cumulative Distribution Function (CDF) for every dimension of each sample across the entire available, continuously streamed dataset towards defining the joint CDF. Thereafter, it computes and scores the propensity of each sample's high-dimensional parameters trending towards an extreme-value relative to the rest of the population, based on its location in the joint CDF. In a multivariate statistical problem, the algorithm is therefore effective and intuitively interpretable in its scoring, while fully accounting for the dataset's multivariate statistical basis. In our problem framing, counts of activation of each step of EM defines each dimension or feature of the model. Consequently, the resulting numerical score output by the model can delineate and rank each memory component and/or SSD as behaving in either "meaningfully insightful" or extrinsic EM behavior, or "median/normal" or intrinsic EM behavior – when accounting for every EM step's activation count in the multivariate joint CDF. In this context, "insightful" or extrinsic is a system-level delineation, such as explained and illustrated in Section V. Such a threshold was found to be effectively and accurately codified as the ECOD model's singular hyperparameter, generalizable across the entire population upon evaluation by storage architects.

Fig. 5 illustrates the ML model-score histogram from each workload class, where the tail of each model's score distribution represents the extrinsic samples, algorithmically identified from among several thousand SSDs and ~0.5 million memory components. As Fig. 5 shows, the scoring needs to be separately ascertained by each workload class. This is because activation counts of EM steps is not merely a function of memory silicon behavior or of EM algorithm behavior – such activation is also reflective of workload stimulus. In other words, some workload classes, due to their specific access patterns at the application software layer can and does activate certain EM steps more, and generally in a different manner, than other workload classes. This fact is reflected in Fig. 5, where the four histograms evidently represent slightly different score distributions depending on the workload class, despite being similarly skewed in nature. Hence, accounting for workload-class differences becomes a necessary aspect of the ML-based modeling methodology.

It is worth noting that the modeling framework was also found amenable to tree-structured ML models, such as Isolation Forest [29]. However, we chose ECOD as our model specifically for its analytically intuitive interpretability alignment with EM and storage architecture design, as strongly preferred by storage system designers and architects. Specifically, directly accessible and interpretable overall model score and feature-wise score of each sample were defining characteristics of the ECOD model's effectiveness as described in Section V. Furthermore, the ECOD model enables Representation Learning of the EM-workload domain, as described Section VI. The following sections delve into details of interpretability illustrated with a specific example and discusses system-level avenues and implications for co-design.

## V. INTERPRETABLE ML MODELING RESULTS

Once extrinsic samples of SSD or memory are identified successfully as represented by a high model score, from the large, heterogeneous and continuously accumulating datasets - the ECOD modeling framework further enables an interpretable understanding relevant to the EM architecture. Above and beyond the identification of insightful samples from comprehensive system design perspectives, causal understanding behind a high score is also highly relevant and necessary for storage architecture design evolution. Towards this goal, Fig. 6 demonstrates the case of an extrinsic sample being identified from the large population undergoing the JEDEC-218/219 class of stress workloads.

This specific illustrative sample had not surfaced any errors at the system level during the entire workload execution. However, from Fig. 6, high scores are consistent for this specific sample from the earliest checkpoint of workload execution, implying the early identification of the sample, despite lacking "fail-stop" or hard errors. From the perspective of storage architects, developing a causal understanding of the high score behind the latent issue, is naturally the next stage of desired insight. With the interpretability inherent in ECOD algorithm design by virtue of being constructed from cumulative distributions for each dimension (i.e. each EM step activation), the model can expose each EM step's individual score underlying the overall high score of the sample. This fact is illustrated in Fig. 6, where the specific sample's scores for each EM step comprising the ECOD model's dimensions, results from directly querying the model. Fig. 6 illustrates the signature of this sample at all steps of EM, relative to the large population. Fig. 6 provides clear evidence, that relative to the $99^{th}$ percentile and 99.9$^{th}$ % percentile scores of the entire population, this specific sample was underperforming the remainder of the large population undergoing the same class of workloads. The sample's excessive need for *deeper* EM steps, more than 99.9 % of the overall population, implies significantly worse, unacceptable latency at the system-level by the host, causing "fail-slow" reliability issues [30]. This actionable causal insight, intuitively understandable from the perspective of storage architecture, leads to deeper analysis of EM algorithms holistically together with the memory silicon, thus contributing to co-design goal. One co-design activity, for example, could be modification of the shallower EM algorithms to more effectively manage the silicon properties – an EM algorithm and firmware consideration that can extend system-level capability. Yet another co-design activity could be silicon design evolution or a screening strategy – a hardware implication. However, each avenue needs evaluation in context of the other, and comprehensive system-level considerations become key. The understanding and enablement of such possibilities are thus fulfilled by the modeling framework.

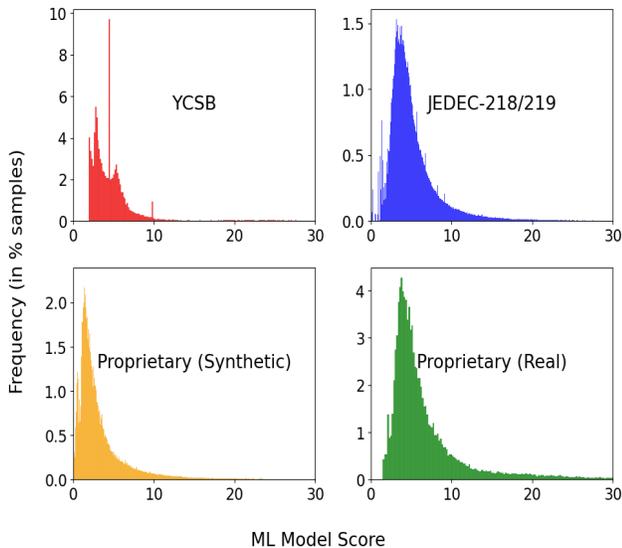

Figure 5. Histograms of multivariate EM-features based ECOD composite model scores, by workload class, for SSD-B as a representative example. Histogram bar are shown as percentages normalized to sum to 100 within each sub-population of workload-classes, to facilitate visualization of the highly right-skewed score distributions.

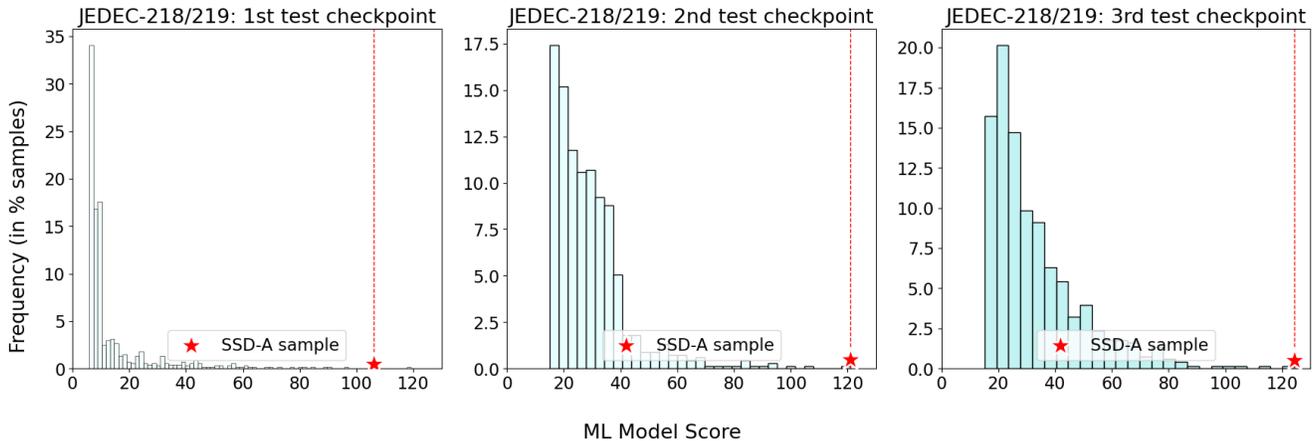

Figure 6. An insightfully meaningful extrinsic sample of SSD-A, consistently identified with a high score from the ECOD multivariate model, across 3 test workload checkpoints, each separated by about a month of time during workload execution. Histogram bars are shown as percentages normalized to sum to 100 to facilitate visualization of the highly skewed score distributions at each checkpoint. The score distributions are seen to evolve across the temporal checkpoints due to continuing EM stimulus generated by workload executions - regardless, the extrinsic sample is consistently surfaced with a high score.

Besides such actionable insights, this modeling on large datasets, for example, also circumvents practical limitations of specialized latency tests and associated hardware being necessarily limited to few SSD samples [31]. Also, while the above description illustrates a concrete real-world case relevant to latency and "fail-slow" reliability for co-design, other diverse areas of both robustness and performance assessment in a granular manner are also achievable with the framework, given the fundamental role of EM design in SSD architecture. Furthermore, workload datasets resulting from specific, targeted designs of experiment for evaluating EM may also be analyzed with this framework - maximally leveraging its granular interpretability, generalizability and reusability. As the ECOD algorithm can be considered an ensemble model in that its aggregate score is an ensemble of scores from each dimension or EM step, the model scales trivially to higher dimensions (i.e. increase in EM step count, $d$) and population size (i.e. increases in SSD population count, $n$) [28]. For the same reason, with the ECOD model having O($nd$) time and space complexity, it is also computationally parallelizable and distributed across commodity multi-core processors -- which eases practical deployment costs without needing accelerators. Thus, we aligned the modeling framework with principles and practical aspects of ML-for-Systems in a deliberate manner [32].

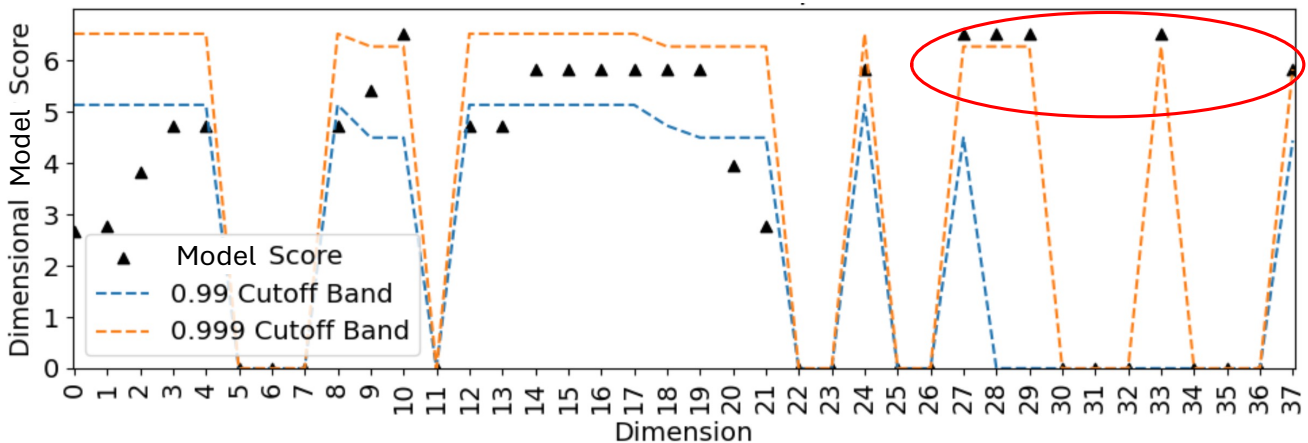

Figure 7. Interpretability of ECOD model enables scoring of each dimension (i.e. EM step), revealing causal reasons behind the *Model Score* of extrinsic sample surfaced in Fig. 6 being significantly worse than 99[th] %-tile and 99.9[th] %-tile of the overall population, for especially the highlighted, deeper EM steps. Such system-level behavior contribute to "fail-slow" or high latency read operations from the storage system, although without any loss of data.

## VI. REPRESENTATION LEARNING ON WORKLOADS

As noted earlier, the datasets of the architected EM stack interacting with workloads encapsulate critical behavioral knowledge associated with robustness, performance and related figures of merit of the storage architecture. When analyzed from perspectives of software applications, we posit that the activation counts of EM steps can allow inferring architectural understanding of their reaction to work-load stimuli through Representation Learning [33]. By leveraging domain-specific knowledge, one can interpret these activation patterns to map the stress profile imposed by workloads - such as dominated by write intensive, read intensive, or environmental factors such as temperature and humidity [3]. Such stimuli can manifest as distinct signatures within the EM steps, enabling classification and comparative analysis of workloads. This interpretive capability can enable storage system architects to:

- Assess variety of workloads and their EM activation profiles as resulting from workload stimuli.
- Identify novel workload patterns and map them along known figures-of-merit (e.g. robustness, performance) spectrum.
- Extend architectural coverage of EM design, enhancing figures-of-merit across diverse operational workload scenarios.

### A. Dimensionality Reduction

In Sections IV & V, direct mapping of every EM step as ML model dimensions or features was crucial for interpretability and co-design goals. However, the 37 or more high-dimensional EM datasets present challenges for direct visualization and interpretability of detailed workload-specific signatures via the numerical scores shown in Fig. 6 and Fig. 7. With our goal described in this section being mapping interpretability of software-layer application signatures via Representation Learning (i.e. going beyond the goal of understanding of the EM architecture for co-design in Sections IV & V), we applied the workflow illustrated in Fig. 8, wherein the ECOD algorithm is first applied *independently* to each EM step across all samples of each workload class. Since the initial, raw EM activations tend to be highly sparse and skewed across samples within each workload, this step ensures continuous distributions such as illustrated in Fig. 5. Thereafter, the median of model scores per EM step is assessed as the *compressed* representation of the reaction of such EM step, across populations of any given workload class. This processing thus yields datasets reflecting stimuli from each workload class on each EM step.

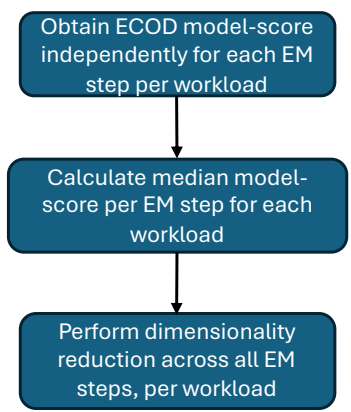

Figure 8. Method for pre-processing datasets for Representation Learning on the domain of EM-workload stimuli.

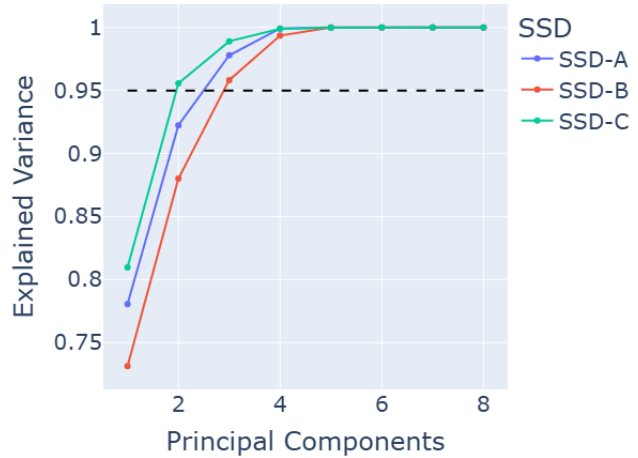

Figure 9. Scree plot of explained variance based on dimensionality reduction with PCA on respective EM steps that vary across different SSD generations.

For subsequent analysis and interpretable visualization, we applied dimensionality reduction on resulting medians of EM step scores, specifically via Principal Component Analysis (PCA) [34]. As shown in the scree plot of Fig. 9, over 95% of variance is explained by three principal components across the population, thereby enabling effective 3-D visualization for compressed workload stimuli representation.

### B. Representation Learning Results

With the dimensionally reduced dataset, we illustrated the visualization of workload behavior in a three-dimensional principal component space in Fig. 10. We found that distinct clustered patterns emerge based on workload similarities, revealing interpretable axes of operational characteristics. Specifically, the axis in red delineates increasing retention stress, the axis in green corresponds to escalating write-intensive workloads, and the axis in purple reflects growing read-intensive workloads. Notably, a cluster comprising mixed read-write workloads is also observed — consistent with expectations from production applications without targeted stresses. Storage system architects can leverage the methodology to enhance resilience or performance by mapping new or unfamiliar workloads onto the established principal component space for EM, thereby informing architectural decisions through such data-driven interpretable workload profiling and modeling.

This avenue of workload-space exploration is expected to lead to research directions that can broaden and deepen the impact of generated insights. As a future research avenue, it may be fruitful to map latency metrics introduced in the Open Compute Project 2.6 specification to the Representation Learning framework, further enhancing co-design opportunities for system-level optimizations [31].

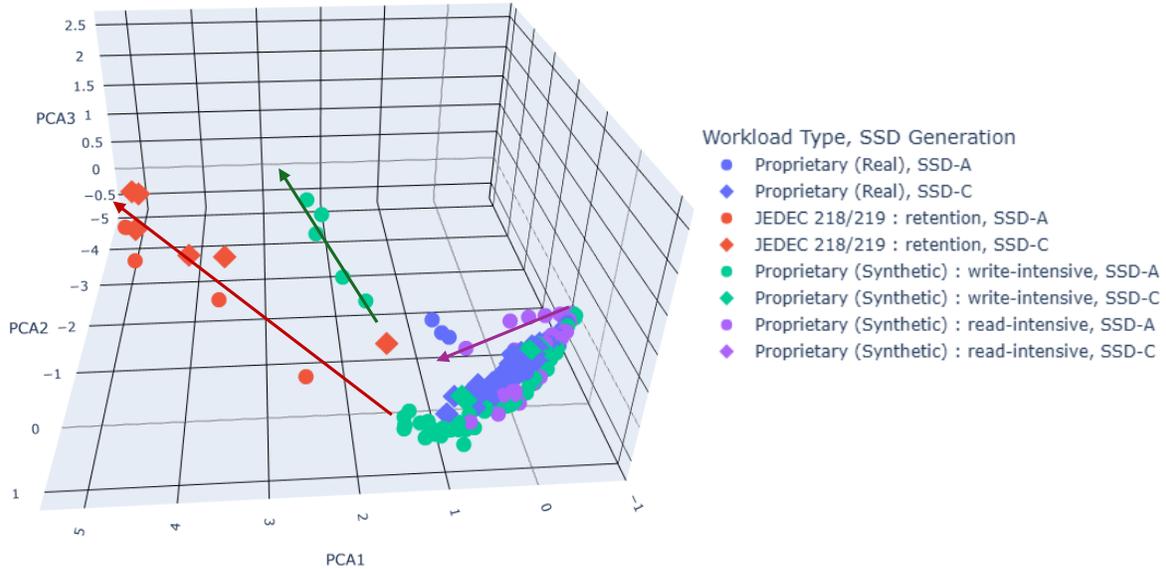

Figure 10. Representation Learning on EM-workload domain across examined SSD generations, where clustered "similarity" representations of EM-workload interaction domain is evident, relevant to workload characteristics with resulting stimulated physics of failure.

## VII. DISCUSSIONS

From the perspective of system-level reliability, the interpretable modeling methodology based on large-scale datasets enabled the dissection of intrinsic versus extrinsic component-level failure modes and mechanisms in a highly granular manner, as illustrated in Section V. Furthermore, causal understanding of the failure mechanisms also offers the possibility of managing such failure mechanisms with novel firmware-enabled algorithms, potentially blurring distinctions between intrinsic and extrinsic silicon mechanisms at system-level, while extending the overall system-level reliability.

While data-driven research approaches and associated deployed ML-for-Systems methodologies are augmented and enhanced by the volume and diversity of the dataset they leverage, they are also inherently limited by a non-infinite volume of the dataset. In concrete terms relevant to our research, the volume and richness of workload classes are equally the enablers and potential limiters of efficacy. The efficacy of our approach is dependent on the completeness of the stimulus spectrum that we expect being captured in the diverse workloads. We acknowledge this critical aspect and enhance mitigation opportunities by relying on the richness of associated datasets. In fields of Natural Language Processing (NLP) or Computer Vision (CV), vast datasets are commodities generally available freely across the internet (setting aside caveats of intellectual property) – thus constituting the entire "basis set" of learnable knowledgebase relevant to NLP and CV. However, in the field of systems research, appropriately rigorous domain knowledge, an "inductive bias", proves necessary together with non-infinite but carefully curated datasets such as we leveraged in our research. We observe that similar fields of scientific ML and technology such as drug discovery and bioinformatics research, where practical risks and costs of incorrect outcomes from applications of ML are also very high, and which also depend on non-infinite but high-value datasets – adopt and argue for similar principles [35], [36]. With that noted, we also take care to not induce unwarranted biases or assumptions in our methodologies, other than ensuring the curation and cleanliness of meaningful datasets together with selecting an interpretable and principled modeling framework.

## VIII. CONCLUSIONS

In our research on storage architecture error management design, we undertook a first-principles, data-driven and ML-for-Systems focused approach that can lead to data-driven storage architecture co-design with holistic system-level considerations. We focused on fully leveraging and entwining domain knowledge of storage architecture with transparent and trusted modeling framework towards optimal benefits both from operational perspectives of co-design and with forward-looking research perspective of evolving workloads and architectures. Specifically, we found and demonstrated that behavioral causal mechanics revealed about the EM stack through data-driven modeling on large-scale datasets can offer rich guidance to storage architects about necessary evolution of either EM algorithms, or silicon design, or both – and offering up the possibility of extending system-level reliability with firmware design. Finally, in defining a Representation Learning space for workloads, we demonstrated a methodology for architectural coverage map that can allow both understanding and accommodating of new and/or changing workloads in the framework for EM architecture codesign for SSD technology.

ACKNOWLEDGMENT

The authors gratefully acknowledge Samir Mittal, Jeff Karpan, Jiangli Zhu, Aaron Lee, Ying Tai, Ramin Ghodsi, Ionicia Dembi and Seth Eichmeyer for their support and feedback on the research.